\documentstyle[prd,aps,psfig]{revtex}
\draft
\newcommand{\be}{\begin{equation}}
\newcommand{\ee}{\end{equation}}
\newcommand{\bn}{\begin{eqnarray}}
\newcommand{\en}{\end{eqnarray}}
\newcommand{\bd}{\begin{displaymath}}
\newcommand{\ed}{\end{displaymath}}
\newcommand{\bnn}{\begin{eqnarray*}}
\newcommand{\enn}{\end{eqnarray*}}
\def\Ref#1{(\ref{#1})}
\def\Journal#1#2#3#4#5{#1,\ {\it #2} \ {\bf #3}, \ #4 \ (#5).}
\def\journal#1#2#3#4#5{#1,\ {\it #2} \ {\bf #3}, \ #4 \ (#5)}
\def\r#1{#1}
\begin{document}
\title{Stochastic Dynamics of Time Correlation 
in Complex Systems with Discrete Time}
\author{Renat Yulmetyev${}^{\dagger\ddagger}$
\thanks{e-mail:rmy@dtp.ksu.ras.ru}, Peter H\"anggi${}^\dagger$, and
Fail Gafarov${}^\ddagger$}  
\address{$\dagger$ Department of Physics,University of Augsburg,
Universit\"atsstrasse 1, D-86135 Augsburg, Germany\\
$\ddagger$\ Department of Theoretical Physics, Kazan State
Pedagogical University,   420021 Kazan, Mezhlauk Street, 1 Russia}
\maketitle
\begin{abstract}
In this paper we present the new concept of
description of random processes in complex systems with the discrete  
time. It involves the description of kinetics of discrete processes 
by means of the chain of finite-difference non-Markov equations for time 
correlation functions (TCF). We have introduced the dynamic (time 
dependent) information Shannon entropy $S_i(t)$ where i=0,1,2,3,...
as an information measure of stochastic dynamics of time correlation
$(i=0)$ and time memory (i=1,2,3,...). The set of functions $S_i(t)$ constitute the 
quantitative measure of time correlation disorder $(i=0)$ and
time memory disorder (i=1,2,3,...) in complex system. The theory developed started
from the careful analysis of time correlation involving 
dynamics of vectors set  of various chaotic states.
We examine two stochastic processes involving
the creation and annihilation of time correlation (or time memory)  
in  details. We carry out the analysis of 
vectors' dynamics employing finite-difference equations
for random variables and the evolution operator
describing  their natural motion. The existence of TCF
results in the construction of the set of projection operators
by the usage of scalar product operation. Harnessing the
infinite set of orthogonal dynamic random variables on a basis
of Gram-Shmidt orthogonalization procedure tends to creation of infinite
chain of finite-difference non-Markov kinetic equations for
discrete TCF and memory functions (MF). The solution of the
equations above thereof
brings to the recurrence relations between the TCF and MF of
senior and junior orders. This offers the new opportunities for 
detecting the frequency spectra of power of entropy function
$S_i(t)$ for time correlation $(i=0)$ and time memory (i=1,2,3,...). The results
obtained offer considerable scope for attack on stochastic dynamics 
of discrete random processes in a complex systems.
Application of this technique on the analysis of stochastic
dynamics of RR-intervals from human ECG's shows convincing
evidence for a non-Markovian phenomemena associated with a
peculiarities in short and long-range scaling. This method may
be of use in distinguishing healthy from pathologic data sets
based in differences in these non-Markovian properties. 
\end{abstract}
\section{Introduction}
Manifold methods are successfully used in
statistical physics for the description of distinctive
characteristics of chaotic dynamics of complex systems
\cite{Li}-\cite{Webber} . Nevertheless,three vexing features 
which are difficult to
yield a detailed and strict analysis are available in complex
systems.  Among them: nonstationarity,
nonlinearity, and nonequlibrium phenomena. Furthermore, the
significant peculiarities of complex systems are directly related
to the discretness of time of object / subject
registration response \cite{Badii}, \cite{Sornette},
\cite{Gaspard}, \cite{Shannon}, \cite{Zebrowski}. Non-Markov and
long-range statistical memory effects also play the leading part in
the complex systems behaviour \cite{Lowen}-\cite{Badii},
\cite{Shannon}, \cite{Fulinski}, \cite{AFul}-\cite{Grifoni}.

However, the discretness of time while considering the
complex systems has not been taken into account until now,
although it is
discretness that is the most commonly encountered feature of real
objects / subjects. On the other hand the memory and time
long-ranging effects are paramount. As a rule the state developed
is complicated by the fact that the real complex
systems are of nonphysical nature.  Therefore, the
direct methods of statistical physics derived from Hamiltonian
formalism, exact equations of motion and Liouville's equation
are not applicable in this case to its theoretical analysis.
Meanwhile the real existence of complex systems in time and 
space generates a reliable evristic basis for the modeling in
terms of the time discretness, memory and time  long-range effects.

The present article is dedicated to statistical consideration of
a discretization in temporary changes of complex systems of a
substantial nature on the basis of the first principles. 
In Section 2 we briefly outline general definitions and proposals
used to form the stochastic dynamics of discrete time sequences,
and in Section 3 we suggest the geometrical presentation of stochastic 
dynamics of time correlation. Introduction of projection operators, 
splitting of equation of states vectors and matrix presentation 
of Liouville's quasioperator for the statistical description of random 
processes with discrete time are reported in Section 4, and  
introduction of the set of orthogonal random variables as well as construction
of infinite chain of finite-difference non-Markov kinetic equations 
for discrete TCF are framed in Section 5. A pseudohydrodynamic 
description of random processes is provided in Section 6, where the
relative merits of this approach are set forth. In Section 7 we 
define Shannon dynamical (time dependent) entropy for time 
correlation and time memory in complex systems. Aplication of
technique on the analysis of stochastic dynamics of RR-intervals
from human ECG's are discussed in Sec. 8. In Section 9 we 
present the discussion and conclusions of the results obtained 
and possible opportunities for the experimental data processing.

\section{Basic assumptions and definitions}
Following Gaspard [15] we consider a random process such as a
sequence of random variables defined at successive times. We
shall denote the random variable by 
\begin{equation}
 X=\{x(T),x(T+\tau),x(T+2\tau),\cdots, 
  x(T+k\tau),\cdots,x(T+\tau N-\tau)\}, \label{f1}
\end{equation} 
which corresponds to signal during the time period $t=(N-1)\tau$
where $\tau$ is time interval of signal discretization.
The mean value $<X>$, fluctuations $\delta x_j$,
absolute $(\sigma^2)$ and relative $(\delta^2)$ dispersion for a
set of random variables \Ref{f1} can be easily found by 
\begin{equation}
<X>=\frac{1}{N}\sum_{j=0}^{N-1} x(T+j\tau), \label{f2}
\end{equation} 
\begin{equation}
 x_j=x(T+j\tau), \delta x_j=x_j-<X>, \label{f3}
\end{equation}
\begin{equation}
\sigma^2=\frac{1}{N} \sum_{j=0}^{N-1} \delta x_j^2, \label{f4}
\end{equation}
\begin{equation}
\delta^2=\frac{\sigma^2}{<X>^2}=\frac{\frac{1}{N}\sum_{j=0}^{N-1}\delta
x_j^2}{\{\frac{1}{N}\sum_{j=0}^{N-1} x(T+j\tau)\}^2}. \label{f5}
\end{equation}
Abovementioned values determine the statical (independent from
time) properties of the system considered. The normalized time
correlation function (TCF) [1-3],[7-9] depending on discrete time
$t=m\tau,N-1 \geq m \geq 1$ can be conveniently used for the
analysis of dynamic properties of complex systems
\begin{equation}
a(t)=\frac{1}{(N-m)\sigma^2} \sum_{j=0}^{N-1-m} \delta
x(T+j\tau) \delta x(T+(j+m\tau)).
\label{f6}
\end{equation}
TCF usage means that developed method is just for complex systems, when
correlation function exist. In forthcoming papers we intend to
apply developed method for discrete random processes analysis in
complex systems in practical psychology, cardiology (for the
development of diagnosis method of cardiovascular diseases),
financial and ecological systems, seismic phenomena, etc.
The properties of TCFs $a(t)$ are easily determined by Eq. \Ref{f6}
\begin{equation}
\lim_{t\to 0} a(t)=1, \lim_{t \to \infty} a(t)=0. \label{f7}
\end{equation}
We have to recognize that the second property in Eq. \Ref{f7}
is not always satistifed for the real systems even with
arbitrary big values of time $t$ or number $(N-1)=t/\tau$.
Taken into account fact that the process is discrete, we
must rearrange all standard operation of
differentiation and integration [34,35] 
\bn
\frac{dx}{dt} \to \frac{\triangle x(t)}{\triangle
t}=\frac{x(t+\tau)-x(t)}{\tau},\   
\int_a^b x(t) dt= \sum_{j=0}^{n-1} x(T_a+j\tau)\triangle t  
=\tau \sum_{j=0}^{n-1} x(T_a+j\tau)=n\tau<X>\ ,\nonumber\\
b-a=c, c=\tau n. \label{f8}
\en
The first derivative on the right is recorded in Eq.\Ref{f8}. The
second derivative on the right is also derived easily 
\be
\frac{d^2 x(t)}{dt^2} \to \frac{\triangle x}{\triangle
t}\left(\frac{\triangle x}{\triangle
t}\right)=\tau^{-2}\{[x(t+2\tau)-x(t+\tau)]
-[x(t+\tau)-x(t)]\}=\tau^{-2}\{x(t+2\tau)-2x(t+\tau)+x(t)\}. 
\label{f9}
\ee
Now let us proceed to the description of the dynamics of the process.
For real systems values $x_j=x(T+j\tau)$ and $\delta
x_j=\delta x(T+j\tau)$  result from the experimental data. Thus
we can introduce in Shannon's manner [17] the 
evolution operator $U(T+t_2,T+t_1)$ in as follows $(t_2\geq t_1)$
\be
 x(T+t_2)=U(T+t_2,T+t_1)x(T+t_1).
\label{f10}
\ee
For brevity let us present Eq.\Ref{f10} in the form 
\be
x(j)=U(j,k)x(k),\   j\geq k, \  j,k=0,1,2,\cdots,N-1. \label{f11}
\ee
Time operator of one step shift $\tau$ along a discrete
trajectory is conviniently considered by means of two nearest values
$x(t+\tau)$ and $x(t)$ 
\be
x(t+\tau)=U(t+\tau,t)x(t). \label{f12}
\ee
Owing to Eqs.\Ref{f10}-\Ref{f12} a formal equation of motion is
derivable for any $x \in (x_0,x_1,x_2,\cdots,x_{N-1})$
\be
  \frac{dx}{dt} \to \frac{\triangle x(t)}{\triangle t}=
\tau^{-1}\{x(t+\tau)-x(t)\}=\tau^{-1}\{U(t+\tau,t)-1\}x(t). \label{f13}
\ee 
Let us consider Eqn.\Ref{f13} in terms of $x_j$ 
\bn
\frac{\triangle x_j(t)}{\triangle
t}=\frac{x_{j+1}(t+\tau)-x_j(t)}{\tau}=\tau^{-1}\{U(t+\tau,t)-1\}x_j(t)
\nonumber 
\en
and then introduce a Liouville's quasioperator $\hat L$ as
follows 
\be
\frac{dx}{dt}=\frac{\triangle x(t)}{\triangle t}=i \hat
L(t,\tau)x(t),  \hat L(t,\tau)=(i\tau)^{-1}[U(t+\tau,t)-1].
\label{f14}
\ee
Now in accordance with Refs. \cite{Gaspard}, \cite{Wehrl} let us 
present a set of values of random
variables $\delta x_j=\delta x(T+j\tau), j=0,1,\cdots,N-1$
as a $k$-component vector of system state 
\be
{\bf A}_k^0(0)=(\delta x_0,\delta x_1,\delta x_2,\cdots,\delta
x_{k-1})=(\delta x(T),\delta x(T+\tau),\cdots,\delta
x(T+(k-1)\tau). \label{f15}
\ee  
Now we can introduce the scalar product operation
\be
  <{\bf A} \cdot {\bf B}>=\sum_{j=0}^{k-1} A_j B_j
  \label{f17}
\ee
with or whithout indication of obvious
time dependence of vectors ${\bf A}$ and ${\bf B}$, respectively,
in the set of vectors ${\bf A}_k^0(0)$ and
${\bf A}_{m+k}^m(t)$ where $t=m\tau$ and
\bn
 {\bf A}_{m+k}^m(t)=\{\delta x_m,\delta x_{m+1},\delta x_{m+2},\cdots,\delta
  x_{m+k-1}\}  \nonumber \\
  =\{\delta x(T+m\tau),\delta x(T+(m+1)\tau), 
  \delta x(T+(m+2)\tau),\cdots,\delta x(T+(m+k-1)\tau\}. \label{f16}
\en
A number $k<N-1$ determines the vectors' dimension. The
functions \Ref{f4}, \Ref{f5} can be expressed in terms of scalar
product \Ref{f17}
\bn
\sigma^2&=&\frac{1}{N}<{\bf A}_N^0 \cdot {\bf
A}_N^0>=N^{-1}\{{\bf A}_N^0\}^2,\nonumber \\
\delta^2&=&\frac{N^{-1}<{\bf A}_N^0 \cdot {\bf A}_N^0>}{<X>^2.} \nonumber
\en 
A $k$- component vector ${\bf A}_{m+k}^m(t)$ displaced to the
distance $t=m\tau$ on the discrete time scale can be formally
presented by the time evolution operator $U(t+\tau,t)$ as follows
\bn
{\bf A}_{m+k}^m(t)=U(T+m\tau,T) {\bf
A}_k^0(0)=\{U(T+m\tau,T+(m-1)\tau)\nonumber \\ U(T+(m-1)\tau,T+(m-2)\tau)\cdots
U(T+\tau,T\} {\bf A}_k^0(0).  \label{f18}
\en
The normalized TCF in Eq. \Ref{f6} can be rewritten in a more compact
form by means of Eqs.\Ref{f16}-\Ref{f18} ( $t=m\tau$ is
discrete time here)
\be
a(t)=\frac{<{\bf A}_k^0 \cdot {\bf A}_{m+k}^m>}{<{\bf A}_k^0 \cdot
{\bf A}_k^0>}=\frac{<{\bf A}_k^0(0) \cdot {\bf A}_{m+k}^m(t)>}{<{\bf A}_k^0(0)^2>}.
\label{f19}
\ee
Replacement of Eqn.\Ref{f6} by Eqn.\Ref{f19} is true if the 
numbers $k<N-1$ satisties the condition
\be
  \sigma^2 \cong k^{-1} \sum_{j=1}^{k-1}\delta x_j^2 ~or~
\sigma^2=\lim_{k \to \infty} k^{-1} \sum_{j=0}^{k-1} \delta
x_j^2. \label{f20}
\ee
 The condition of quasi-stationarity of processes under consideration
\be
\left|\frac{da(T,t)}{dT}\right| \ll
\left|\frac{da(t)}{dt}\right|, \label{f21}
\ee
serves the other criterion of validity for such replacement.
TCF $a(T,t)$ at Eqn.\Ref{f21} is viewed on a time scale 
(point $T$) at the distance $t$ from the zero point .

Such vectors' notion is very helpful for the
analysis of dynamics of random processes by means of
finite-difference kinetic equations of non-Markov type.
\section{Geometrical notion of stochastic dynamics of time
correlation} 
First of all let us consider the projection operation in the set of
vectors for different system states. It is easy to introduce it employing
the above scalar product \Ref{f17}. Then it is necessary to
introduce vectors ${\bf A}= {\bf
A}^0_k(0)$ and ${\bf B}={\bf A}^m_{m+k}(m \tau)$. Using
Fig.1 and simple geometrical notions  
we can demonstrate the following relations in terms of these symbols
\bn
&1)&<{\bf A} \cdot {\bf B}>=|{\bf A}|\cdot |{\bf B}| \cos
\vartheta, \ \cos\vartheta=a(t), \nonumber \\
&2)&{\bf B}={\bf B}_\parallel+{\bf B}_\perp;\nonumber \\
&3)&{\bf B}_\parallel=|{\bf B}| \cos \vartheta \frac{{\bf
A}}{|{\bf A}|}=\frac{{\bf A}}{|{\bf A}|}|{\bf B}| a(t), |{\bf
B}_\parallel|^2=|{\bf A}|^2\{a(t)\}^2, \nonumber \\
&4)& |{\bf B}_\perp|=|{\bf B}| \sin \vartheta=|{\bf
B}|\{1-[a(t)]\}^{1/2}, \label{f22}
\en
where symbol $|{\bf A}|$ denotes the vector ${\bf A}$ length.
Geometrical distance $R({\bf A}, {\bf
B})$ between two vectors ${\bf A}$ and ${\bf B}$ can also be found
\bn 
R({\bf A},{\bf
B})=\{|{\bf A}-{\bf B}|^2\}^{1/2}=\{\sum_{j=0}^{k-1}({\bf A}_j-{\bf
B}_j)^2\}^{1/2}. \nonumber
\en 
Using the latter and taking into account Eqs. \Ref{f6},
\Ref{f19}  we can find 
\bn
R({\bf
A}_k^0(0),{\bf A}_{m+k}^m(t))=\{|{\bf
A}_{m+k,\perp}^m(t)|^2\}^{1/2}=
\sqrt{2}|{\bf A}_{m+k}^m(t)|\{1-a(t)\}^{1/2}. \nonumber
\en
The equation above immediately shows that the distance is determined by
the dynamics of evolution of correlation process. Owing to the
property \Ref{f7} the following relation 
 $\lim_{t\to \infty} R({\bf A}_k^0(0)),
{\bf A}_{m+k}^m(t))=\sqrt{2 k\sigma^2}$, where $\sigma^2$ is the
variance can be developed. With regard to Eqn. \Ref{f22} the
correlation decay in limit $t\to\infty$ may result in complete
annihilation of parallel component of state  ${\bf  
A}_{m+k}^m(t)$ vector . Then the state of the system at the moment $t \to \infty$ 
is entirely determined by the perpendicular component ${\bf A}_{m+k,\perp}^m(t)$
of the full vector ${\bf A}_{m+k}^m(t)$.

It follows from Eqs. \Ref{f22}  that in the set of 
state $\{{\bf A}_k^0(0), {\bf A}_{m+k}^m(t)\}$ vectors at
different values of $t,~m$ and $k$, TCF of random processes
$a(t)$ plays a crucial role as an indicator of two interrelated states
of a complex system. One of them deals with the creation
of correlation and is specified by the ${\bf B}_\parallel$
component, whereas the second one is related to the annihilation
of correlation and determined by the component ${\bf B}_\perp$.
It results in the fact that in the limit of great $t \to \infty$
the following relation: \be
  \lim_{t\to \infty} {\bf A}_{m+k,\parallel}^m(t)=0,~~~
\lim_{t\to \infty} {\bf A}_{m+k,\perp}^m(t)={\bf
A}_{m+k}^m(t). \label{f23}
\ee  
is immediately fulfilled in correspondence with to Bogolubov's 
\cite{Bogolubov} principle of correlation attenuation.

From the physical point of view this fact means that TCF
$a(t)$ represents two interrelated states determined by 
creation and annihilation of correlation. Hence it follows that
such consideration must be given to both processes in an
explicit form for stochastic dynamics  of random
processes' correlation. 
\section{Splitting of equation  of vectors motion and
Liouvillian's matrix presentation}
It is obvious from Eq.(22) that TCF $a(t)$ is
originated by projection of vector ${\bf A}_{m+k}^m(t)$
\Ref{f18}, where time $t=m\tau$ on the initial vector of state
${\bf A}_k^0(0)$ (see, for example, formula \Ref{f19}).  The 
following construction of projection operator 
\be
\Pi {\bf A}_{m+k}^m(t)={\bf A}_k^0(0)\frac{<{\bf A}_k^0(0) {\bf
A}_{m+k}^m(t)>}{<|{\bf A}_k^0(0)|^2>}={\bf A}_k^0(0) a(t) \label{f24}
\ee
results from here .
It is turn projection operator $\Pi$ from Eqn. \Ref{f24} 
has the following properties
\be
\Pi=\frac{{|\bf A}_k^0(0)><{\bf A}_k^0(0)|}{<|{\bf
A}_k^0(0)|^2>},~~ \Pi^2=\Pi, ~~P=1-\Pi,~~P^2=P,~~\Pi P=0, P\Pi=0. \label{f25}
\ee 
A pair of projection operators $\Pi$ and $P$ are idempotent and mutually 
supplementary. Figure 1 shows that projector
$\Pi$ projects on the direction ${\bf A}_k^0(0)$, whereas the
orthogonal operator $P$ transfers all vectors to the orthogonal
direction. 

Let us consider quasidynamic finite-difference Liouville's Eq.
\Ref{f14} for the vector of fluctuations 
\be
\frac{\Delta}{\Delta t}{\bf A}_{m+k}^m(t)=i \hat L(t,\tau){\bf
A}_{m+k}^m(t).  \label{f26}
\ee
The vectors ${\bf A}_{m+k}^m(t)$ generate the vector finite-dimensional
space $A(k)$ with scalar product
 in which (according to Eqn.\Ref{f24}-\Ref{f25}) the
 orthogonal projection operation
\cite{Ljusternik},\cite{Grabet} is expressed by
\bn
  A(k)=A'(k)+A''(k),~~{\bf A}_{m+k}^m(t) \in A(k), \nonumber \\ 
  A'(k)=\Pi A(k), ~~A''(k)=P A(k)=(1-\Pi)A(k). \label{f27}
\en
Operators $\Pi$ and $P$ split Euclidean space  $A(k)$ into
two mutually-orthogonal subspaces. This permits to split
dynamical equation \Ref{f26} into two equations within two
mutually-supplementary subspaces \cite{Grabet}-\cite{Yulmety} as
follows
\bn
\frac{\Delta A'(t)}{\Delta t}=i \hat L_{11} A'(t)+i \hat L_{12}
A''(t), \label{f28} \\
\frac{\Delta A''(t)}{\Delta t}=i \hat L_{21} A'(t)+i \hat L_{22}
A''(t).\label{f29}
\en
In the Eqs. above we crossout for short space elements indices
$A, A'$ and $A''$ and matrix elements arguments $\hat L_{ij}$,
$\hat L_{ij}=\Pi_i \hat L
\Pi_j,~~\Pi_1=\Pi,~~\Pi_2=P=1-\Pi,~~i=1,2.$ 
In line with Refs.\cite{Yulmetyev}, \cite{Yulmety} we write down
Liouville's operator in matrix form 
\bn
\hat L=\left(\begin{array}{cc}
        \hat L_{11} & \hat L_{12}\\ 
        \hat L_{21} & \hat L_{22}
        \end{array}\right), \nonumber \\
\hat L_{11}=\Pi \hat L \Pi,~~\hat L_{12}=\Pi \hat L P, \nonumber \\ 
\hat L_{21}=P \hat L \Pi,~~\hat L_{22}=P \hat L P. \label{f30}  
\en
Operators $\hat L_{ij}$ act in the  following way: $\hat
L_{11}$ - from $A'$ to $A'$, $\hat L_{22}$ - from $A''$ to
$A''$, $\hat L_{21}$ - from $A'$ to $A''$, and $\hat L_{12}$
operates from $A''$ to  $A'$.

To simplify Liouville's Eq. \Ref{f28},
\Ref{f29} we exclude the irrelevant part $A''(t)$ and
construct closed Equation for relevant part $A'(t)$. For
this purpose let us solve Eqn.\Ref{f29}  step by
step 
\be
\frac{\Delta}{\Delta t}\{{\bf A}_{m+k}^m(t)\}''=i\hat L_{21}\{{\bf
A}_{m+k}^m(t)\}' +i\hat L_{22}\{{\bf A}_{m+k}^m(t)\}''.\label{f31}
\ee
Considering Eqn.\Ref{f8} we arrive at finite-difference solution of
this equation in the following form
\bn
\frac{\Delta A''(t)}{\tau}=\tau^{-1}[A''(t+\tau)-A''(t)]=i\hat
L_{21} A'(t)+i\hat L_{22} A''(t), \label{f32} \\
A''(t+\tau)=\{1+i \tau \hat L_{22} \}A''(t)+i\tau \hat
L_{21}A'(t). \label{f33}
\en 
Applying Eqs.\Ref{f32}, \Ref{f33} we find 
\be
A''(t+2\tau)=(1+i\tau \hat L_{22})^2 A''(t)+(1+i\tau \hat
L_{22})\{ i\tau \hat L_{21} A'(t)\}+\{ i\tau \hat L_{21}\}A'(t+\tau),\label{f34}
\ee
for $m=2$ and

\be
A''(t+3\tau)=(1+i\tau \hat L_{22})^3 A''(t)+(1+i\tau \hat
L_{22})^2 \{i\tau \hat L_{21} A'(t)\}+(1+i\tau \hat
L_{22}) \{i\tau \hat L_{21} A'(t+\tau)\}+\{i \tau \hat L_{21}
A'(t+2\tau \},\cdots~, \label{f35}
\ee
for $m=3$, respectively.
In general case  we find
\be
A''(t+m\tau)=\{1+i \tau \hat L_{22}\}^m
A''(t)+\sum_{j=0}^{m-1}\{ 1+i \tau \hat L_{22}\}^j\{ i \hat
L_{21} A'(t+(m-1-j)\tau)\} \label{f36}
\ee
for the arbitrary number of  $m$-steps.
Then after the substitution of right side of
Eq.\Ref{f36} for Eq.\Ref{f28} we obtain the closed 
finite-difference kinetic equation for the relevant parts of vectors 
\be
\frac{\Delta}{\Delta t}A'(t+m\tau)=i \hat L_{11}
A'(t+m\tau)+i\hat L_{12}\{1+i\tau \hat L_{22}\}^m A''(t)- \hat
L_{12}\sum_{j=0}^{m+1}\{1+i\tau \hat L_{22}\}^j\tau \hat
L_{21}A'(t+(m-1-j)\tau). \label{f37}
\ee
To simplify this equation, let us consider the idempotentity
property, and then determine $(0\leq k\leq m-1)$ 
\be
A''(t)=0,~~\{1+i\tau \hat L_{22}\}^k A''(t)=0.\label{f38}
\ee
Transfering from vectors ${\bf A}_{m+k}^m$ in Eq.\Ref{f37} to a
scalar value of TCF $a(t)$ by means of suitable projection we come
to the closed finite-difference discrete equation for the initial TCF
\be
\frac{\Delta a(t)}{\Delta t}=i \omega^{(0)}_0 a(t)-\tau
\Omega_0^{2} \sum_{j=0}^{m-1} M_1(j\tau) a(t-j\tau).\label{f39}
\ee
Here $\Omega_{0}$ is the general relaxation frequency whereas
frequency $\omega_0^{(0)}$ describes the eigenspectrum of the
Liouville's quasioperator $\hat L$
\be
\omega_0^{(0)}=\frac{<{\bf A}_k^0(0) \hat L
{\bf A}_k^0(0)>}{<|{\bf A}_k^0(0)|^2>},~~ \Omega_0^2=\frac{<{\bf
A}_k^0 \hat L_{12}\hat L_{21} {\bf A}_k^0(0)>}{<|{\bf A}_k^0(0)|^2>}.\label{f40}
\ee
Function $M_1(j\tau)$ in the right side of Eq.(39) is the first
order memory function
\be
M_1(j\tau)=\frac{<{\bf A}_k^0(0) \hat L_{12}\{1+i\tau \hat
L_{22}\}^j \hat L_{21} {\bf A}_k^0(0)>}{<{\bf A}_k^0(0) \hat
L_{12} \hat L_{21} {\bf A}_k^0(0)>},~~M_1(0)=1.\label{f41}
\ee
Equation (39) alongside with Eqs.(40),(41) present first order discrete
non-Markov kinetic equation for the discrete time correlation
function $a(t)$. However, our consequent step will be to perform
a further generalization of discrete TCF analysis
and to obtain finite-difference equation for the first order
memory function
$M_1(j\tau)$ and so on.
\section{Introduction of the set of orthogonal random variables and
construction of infinite chain of finite-difference non-Markov kinetic equations
for discrete Memory Functions}
The discrete memory function $M_1(j\tau)$ \Ref{f41} in Eq.
\Ref{f29} is in its turn the normalized TCF, evolution of which is
defined by the deformed (compressed) Liouvillian's $(\hat L^{(0)}=\hat L)$
\be
\hat L^{(1)}=\hat L_{22}^{(0)}=\hat L_{22}=(1-\Pi)\hat L (1-\Pi)
\label{f42}
\ee
for a new dynamical variable $B^{(1)}=i \hat L_{21}{\bf
A}_k^0(0)$. Thus, we can completely repeat for $M_1(j \tau)$ the whole
procedure within Eqs. \Ref{f24}-\Ref{f41}, and
obtain the following non-Markov kinetic equation for the normalized TCF. The
infinite chain of equations for the initial TCF and memory
functions of increasing order results from multiple
repetition of similar procedure. 

However this chain of equations can be obtained differently,
i.e. much shorter and less costly. For this purpose let us employ the method
developed earlier  for the physical
Hamilton systems with the continuous time in Refs.
\cite{Yulmety}, \cite{Yulmet}. Moreover the lack of
Hamiltonian and the time discretness must be taken into account.

Let us remember that natural equation of motion \Ref{f14}
is the finite-difference Liouville's equation 
\be
   \frac{\Delta}{\Delta t}x(t)=i \hat L x(t) \label{f43}
\ee
where Liouville's quasioperator is
\be
     \hat L=\hat L(t,\tau)=(i\tau)^{-1}\{U(t+\tau,t)-1\}. \label{f44}
\ee
 Succesively applying the quasioperator $\hat L$ to the dynamic variables
${\bf A}_{m+k}^m(t)$ ($t=m\tau$, where $\tau$ is a discrete time
step) we obtain the infinite set of 
dynamic functions
\be
  {\bf B}_n(0)=\{ \hat L\}^n {\bf A}_k^0(0),~~n\geq 1. \label{f45}
\ee

Using variables ${\bf B}_n(0)$ one can find the formal solution
of evolution Eq.(43) in the form
\be
{\bf A}_{m+k}^m(m\tau)=\{1+i\tau \hat L\}^m {\bf
A}_k^0(0)=\sum_{j=0}^m \frac{m!(i\tau)^{m-j}}{j!(m-j)!}{\bf
B}_{m-j}^0(0). \label{f46}
\ee
However, a similar form of dynamic variables is deficient. 
That is why we prefer the use the orthogonal
variables as vectors ${\bf W}_n$ given below. 
  Employing Gram-Schmidt orthogonalization procedure [42] for the set 
of variables ${\bf B}_n(0)$ one can obtain the new infinite set of dynamical 
orthogonal variables, i.e. vectors ${\bf W}_n$
\be
 <{\bf W}_n^*(0),{\bf W}_m(0)>=\delta_{n,m}<|{\bf W}_n(0)|^2>, \label{f47}
\ee
where the mean  $<\cdots>$ should be read in terms of  Eqs. \Ref{f17}-\Ref{f18} and $\delta_{n,m}$
is Kronecker's symbol.
  Now we may easily introduce the recurrence formula in which the senior
values ${\bf W}_n={\bf W}_n(t)$ are connected with the juniour
values 
\be
{\bf W}_0={\bf A}_k^0(0),~~{\bf W}_1=\{\hat
L-\omega_0^{(0)}\}{\bf W}_0,~~{\bf W}_n=\{\hat
L-\omega_0^{(n-1)}\}{\bf W}_{n-1}-\Omega_{n-1}^2 W_{n-2},~~n>1.
\label{f48} 
\ee
   Here we used the equation, given earlier in (40) for number
n=0 
\be
\omega_0^{(n)}=\frac{<{\bf W}_n \hat L {\bf W}_n>}{<|{\bf
W}_n|^2>},~~ \Omega_n^2=\frac{<|{\bf W}_n|^2>}{<|{\bf W}_{n-1}|^2>},  \label{f49}
\ee
where $\Omega_n$ is the general relaxation frequency, and 
frequency $\omega_0^{(n)}$ completely describes the eigen 
spectrum of Liouville's quasioperator $\hat L$  \Ref{f44} .
     Now the arbitrary variables ${\bf W}_n$ may be expressed directly 
through the inital variable ${\bf W}_0={\bf A}_k^0(0)$ by means of Eqs. (48)
\bn
{\bf W}_n=\left| \begin{array}{ccccc}
\hat L-\omega_0^{(0)} & \Omega_1 & 0 & \ldots & 0 \\ 
\Omega_1 & \hat L-\omega_0^{(1)} & \Omega_2 & \ldots & 0\\
0 & \Omega_2 &   \hat L-\omega_0^{(2)} & \ldots & 0 \\
0 & 0 & 0 & \ldots &  \hat L-\omega_0^{(n-1)}\\
\end{array} \right| {\bf W}_0. \label{f50}
\en
The physical sense of ${\bf W}_n$ variables (vectors of state)
can be cleared up
in the following way. For example, in the continuous matter
physics, the local density 
fluctuations may be considered as initial variables.
So the local flow density, energy density and energy flow density 
fluctuations are the dynamic variables ${\bf W}_n$ where numbers $n \geq 1$.
The careful usage of the abovementioned variables within the long-wave limits 
creates the basis for the condensed matter theory in hydrodynamic
approximation.
  The set of the orthogonal variables (48) [see also (47)] can be 
connected with the set of projection operators.The later projects the
arbitrary dynamic variable (i.e.,vector of state) $Y$ on the corresponding 
vector of the set
\bn
\Pi_n=\frac{| {\bf W}_n><{\bf W}_n^*|}{<|{\bf
W}_n|^2>},~~ \Pi_n^2=\Pi_n,~~ P_n=1-\Pi_n,~~ P_n^2=P_n,~~\Pi_n
P_n=0,\nonumber \\
\Pi_n\Pi_m=\delta_{n,m} \Pi_n,~~P_n P_m=\delta_{n,m}
P_n,~~ P_n\Pi_n=0.  \label{f51}
\en
    Let us take into consideration the fact that both sets (45) and (50) are infinite.
If we execute the operations in the Euclidean space of dynamic variables then
the formal expressions (51) must be understood as follows
\be
\Pi_n{\bf Y}={\bf W}_n \frac{<{\bf W}_n^* {\bf Y}>}{<|{\bf
W}_n|^2>},~~ {\bf Y}\Pi_n={\bf W}_n^*\frac{<{\bf Y} W_n>}{<|{\bf
W}_n|^2>}.   \label{f52}
\ee
     Now according to Eqs. (28)-(30), (51),(52) we can introduce the 
following notation for the splitting of the Liouville's quasioperator
into the diagonal $(\hat L_{ii}^{(n)})$ and non-diagonal $(\hat
L_{ij}^{(n)})$ matrix elements with $i \not= j,~~n \geq 1$
\bn
\hat L^{(n)}=P_{n-1}\hat L^{(n-1)}P_{n-1},~~\hat L_0=\hat L,
\hat L_{ij}^{(n)}=\Pi_i^{(n-1)} \hat L \Pi_j^{(n-1)},~~i,j=1,2,
\nonumber \\
\Pi_1^{(n)}=\Pi_n,~~\Pi_2^{(n)}=P_n=1-\Pi_n.                                           
 \label{f53}
\en
 For example,  we come to the following Eqs.
\be
\hat L_{22}^{(0)}=\hat L_0=\hat L,~~\hat
L_{22}^{(n)}=P_{n-1}P_{n-2} \ldots P_0 \hat L P_0 \ldots
P_{n-2}P_{n-1} . \label{f54} 
\ee
for the second diagonal matrix elements.
  Successively applying projection operators $\Pi_n$ and $P_n$ for the discrete
equation (43) in the set of normalized TCF ($t=m\tau$)
\be
 M_n(t)=\frac{<{\bf W}_n[1+i\tau \hat L_{22}^{(n)}]^m {\bf
 W}_n>}{<|{\bf W}_n(0)|^2>} \label{f55} 
\ee
we obtain the infinite hierarchy of connected non-Markov 
finite-difference kinetic equations ($t=m\tau$)
\be 
\frac{\Delta M_n(t)}{\Delta t}=i \omega_0^{(n)} M_n(t)-\tau
\Omega_{n+1}^2 \sum_{j=0}^{m-1} M_{n+1}(j\tau)M_n(t-j\tau), \\
 \label{f56}
\ee
where $\omega_0^{(n)}$ is the eigen and $\Omega_n$ is the
general relaxation frequency as follows
\bn
\omega_0^{(n)}=\frac{<W_n^* L_n W_n>}{<|W_n|^2>},
 \ L_n=L_{22}^{(n)}, \ \Omega_n^2=\frac{<|W_n|^2>}{<|W_{n-1}|^2>}.
\nonumber
\en
 A set of functions $M_n(t)$ (55), (56) except $n=0$
\bn 
M_0(t)=a(t)=\frac{<{\bf A}_k^0(0) {\bf A}_{m+k}^m(t)>}{<|{\bf
A}_k^0(0)|^2>}, ~~t=m\tau     \nonumber
\en
  can be considered as functions characterizing the statistical 
memory of time correlation in the complex systems with discrete 
time.
  The initial TCF a(t)  and the set of discrete memory
functions $M_n(t)$ in Eq. (56) are of crucial role for the further consideration. It is 
convenient to rewrite the set of discrete kinetic Eqs.(56) as the 
infinite chain of coupled non-Markov discrete equations of nonlinear
type for the initial discrete TCF a(t) ( discrete time
$t=m\tau$ everywhere)
\bn
\frac{\Delta a(t)}{\Delta t}&=-&\tau \Omega_1^2 \sum_{j=0}^{m-1}
M_1(j\tau) a(t-j\tau)+i\omega_0^{(0)}a(t), \nonumber \\
\frac{\Delta M_1(t)}{\Delta t}&=-&\tau \Omega_2^2 \sum_{j=0}^{m-1}
M_2(j\tau) M_1(t-j\tau)+i\omega_0^{(1)}M_1(t), \nonumber \\
\frac{\Delta M_2(t)}{\Delta t}&=-&\tau \Omega_3^2 \sum_{j=0}^{m-1}
M_3(j\tau) M_2(t-j\tau)+i\omega_0^{(2)}M_2(t).
\label{f57}
\en
These finite-difference Eqs. \Ref{f56} and \Ref{f57}
are very similar to famous Zwanzig'-Mori's chain (ZMC) of kinetic equations
\cite{Zwanzig}- \cite{Evans},
which plays the fundamental role in modern statistical physics of
nonequilibrium phenomena with the smooth time. It should be
noted that ZMC'c is true only for the physical quantum
and classical systems with smooth time governed by Hamiltonian.
 Our finite-difference kinetic equations \Ref{f56}, 
\Ref{f57} are valid for complex systems lacking Hamiltonian,
the time being discrete and the exact equations of motion being absent.
However, the "dynamics" and "motion" in the real complex systems
are undoubtedly abundant and are immediately registered during
the experiment. 

The first three of those Eqs.\Ref{f57} in the whole infinite chain
\Ref{f56} form the basis for the quasihydrodynamic description 
of random processes in complex systems.
\section{A pseudohydrodynamic description of random processes in complex
systems}
At first let's find the matrix elements $\hat L_{ij}$
 of complex systems Liouvillian's quasioperator. Employing 
Eqs. \Ref{f24}, \Ref{f25},\Ref{f30},\Ref{f53} and \Ref{f54} we successively
found  
\be
i\hat L_{11}^{(0)}=\Pi \frac{a(\tau)-a(0)}{\tau}=a'(0)\Pi, \label{f58}
\ee
\be
i\hat L_{21}^{(0)}=\{\tau^{-1}[U(t+\tau,t)-1]-a'(0)\}\Pi,  \label{f59}
\ee
\be
i\hat L_{12}^{(0)}=\Pi\{\tau^{-1}[U(t+\tau,t)-1]-a'(0)\},  \label{f60}
\ee
\bn
i\hat L_{22}^{(0)}=i\hat L-i\{\hat L_{11}^{(0)}+\hat
L_{12}^{(0)}+\hat
L_{21}^{(0)}\}=\tau^{-1}[U(t+\tau,t)-1]-\nonumber \\
\tau^{-1}\Pi\{U(t+\tau,t)-1\}-
\tau{^-1}\{U(t+\tau,t)-1\}\Pi+a'(0)\Pi.      \label{f61}
\en
A diagonal matrix element $\hat L_{22}^{(0)}$ is the part of
"compressed" evolution quasioperator, which in its turn is equal
to
\be
1+i\tau \hat L_{22}=U(t+\tau,t)+\tau a'(0) \Pi-\{\Pi,U(t+\tau,t)-1\}_+,  \label{f62}
\ee
where the anticommutator of appropriate operator is designate by the
brackets $\{A,B\}_+=AB+BA$.One can see from the Eq.\Ref{f62}
that the "compressed" evolution operator differs from the natural
operator $U(t+\tau,t)$ because of the presence of contributions,
associated with the first and the following derivatives of TCF 
the initial TCF $a(t)$.

The large-scale presentation of the memory function $M_1(t)$ is
suitable mostly for practical applications. Using 
Eqs. \Ref{f58}-\Ref{f61}, and \Ref{f41}, \Ref{f54} we also find
the succession of the first five points of discrete functions
$M_1(j \tau)$ where $j=1,2,3,4$ and
\be
M_1(0)=1, M_1(\tau)=\{1/a''(0)\}\{a''(\tau)-2\tau a'(0) a''(0)\}.  \label{f63}
\ee
The "Gaussian" behavior of TCF at the zero point $t=0$
\bn
a'(0)=\frac{<{\bf A}_k^{(0)}(0)\{U_\tau-1\}{\bf
A}_k^{(0)}(0)>}{<|{\bf A}_k^0(0)|^2>}=0  \nonumber
\en
should be taken into account in the subsequent discussion.
It is proved accurately and connected directly with the
orthogonality property of dynamical variables
\Ref{f47}-\Ref{f50}. It gives us an opportinity to simplify the
memory function formula as follows
\bn
M_1(0)&=&1,~~M_1(\tau)=\{\frac{a''\tau}{a''(0)}\},\nonumber \\
M_1(2\tau)&=&\{\frac{1}{a''(0)}\}\{a''(2\tau)-2\tau
a''(0)a'(\tau)+\tau [a''(0)]^2\}, \nonumber \\
M_1(3\tau)&=&\{\frac{1}{a''(0)}\}\{a''(3\tau)-\tau
a'(2\tau)a''(0)-2\tau a''(\tau) a'(\tau)+\tau^2
a''(\tau)a''(0)+\tau a'(\tau)a''(0)\}, \nonumber \\
M_1(4\tau)&=&\{\frac{1}{a''(0)}\}\{a''(4\tau)-\tau
a'(3\tau)a''(0)-\tau a'(2\tau)a''(\tau)+\tau a'(2 \tau
)a''(0)- \nonumber \\
\tau a''&(\tau)& a'(2\tau)+\tau^2[a''(\tau)]^2-\tau a''(\tau)
a'(\tau)+ \tau^2 a''(\tau) a(\tau)a''(0)-\tau^2 a(\tau)[a''(0)]^2\}.
   \label{f64}
\en
Further presentation the following values $M_i(j\tau)$ 
(numbers $j\geq 5$) constitute the
extremely complicated combinatory problem. As analysis of
Eqs.\Ref{f64} shows second derivative's behaviour 
\be
M_1(j \tau ) \cong \{\frac{1}{a''(0)}\}a''(j\tau)   \label{f65}
\ee
 contributes mainly into functions $M_1(j\tau)$.  

Now let us move to practical realization of Eqs.\Ref{f57},
forming a basis of  pseudohydrodynamic description of
correlation dynamics. Thus using orthogonal dynamic variables
\Ref{f47}, \Ref{f48},\Ref{f50}, we immediately obtain 
\bn
\hat W_0= {\bf A}_k^{0},~~\hat W_1=\{\hat L-\omega_0^{(0)}\} \hat
W_0=\hat L \hat W_0=(i\tau)^{-1}(U_\tau-1){\bf A}_k^0(0), \nonumber\\
\hat W_2=\hat L \hat W_1-\Omega_1^2 \hat W_0=\{\hat
L^2-\Omega_1^2\} \hat W_0=(i\tau)^{-2}\{U_\tau-1\}^2{\bf
A}_k-\Omega_1^2 {\bf A}_k^0,\nonumber \\
\hat W_3=\hat L \hat W_2-\Omega_2^2 \hat W_1=\hat L(\hat
L^2-\Omega_1^2) \hat W_0-\Omega_2^2 \hat L \hat W_0= \nonumber
\\ \{\hat L^3-(\Omega_1^2+\Omega_2^2)\hat L\} \hat
W_0=\{(i\tau)^3[U_\tau-1]^3-(i\tau)^{-1}(\Omega_1^2+\Omega_2^2)(U_\tau-1)\}
{\bf A}_k^0.
   \label {f66}
\en
 Simple relation for the eigen and general relaxation frequencies
\bn
\omega_0^{(n)}=\frac{<\hat W_n \hat L \hat W_n>}{<|\hat
W_n|^2>}=0,~~\Omega_n^2=\frac{<|\hat W_n|^2>}{<|\hat W_{n-1}|^2>},
\Omega_1^2=|a^{(2)}(0)|, \nonumber \\
~~\Omega_2^2=\frac{a^{(4)}(0)-
(a^{(2)}(0))^2}{|a^2(0)|} ,~~
\Omega_3^2=\frac{a^{(6)}(0)-2a^{(4)}(0)(\Omega_1^2+\Omega_2^2)-(\Omega_1^2+
\Omega_2^2)^2 a^{(2)}(0)}{a^{(4)}(0)-(a^{(2)}(0))^2}
 \label{f67}
\en
should be taken into consideration here.
The orthogonal variables $\hat W_n$ in Eq.\Ref{f66} can be easily
rearranged as follows
\bn
\hat W_0={\bf A}_k^0,~~\hat W_1=-i\frac{d}{dt}{\bf
A}_k^0=-i\frac{\Delta}{\Delta t}{\bf A}_k,\nonumber \\
\hat W_2=\{\frac{d^2}{dt^2}+\Omega_1^2\} {\bf
A}_k^{(0)}=\{\left(\frac {\Delta}{\Delta
t}\right)^2+\Omega_1^2\}{\bf A}_k^0,\nonumber \\
\hat W_3=i\left\{\frac{d^3}{dt^3}+(\Omega_1^2+\Omega_2^2)\frac{d}{dt}\right\}
{\bf A}_k^0=i\left\{ \left(\frac{\Delta}{\Delta
t}\right)^3+(\Omega_1^2 +\Omega_2^2)\frac{\Delta}{\Delta t}\right\}{\bf A}_k^0.
\label{f68}
\en
Those formulas \Ref{f68} have considerable utility inasmuch  as they
permit to see the structure of formation of orthogonal variables
and junior orders memory functions for the numbers $n=1,2,3$. 
Equations \Ref{f64}, \Ref{f67}, \Ref{f68} open up new fields of
construction of quasikinetic description of random processes $\{{\bf
A}_k^0(0),{\bf A}_{m+k}^{m}(m\tau)\}$. 
By analogy with hydrodynamics the variables $\hat W_0,~\hat W_1,
~\hat W_2$ and  $\hat W_3$ 
in Eq. \Ref{f68} play the role similar to that of the local density,
local flow, local energy density and energy flow. 
It is clear that this is only formal analogy and
the variables $\hat W_n$ dont possess any physical
sense. However, such analogies can be helpful in revealing of
the real sense of orthogonal variables.

To describe pseudohydrodynamics we have to use the set
of first three discrete kinetic Eqs. \Ref{f57} with frequences
$\Omega_i^2~~(i=1,2,3)$ derived from Eqs.\Ref{f67}. It is essential 
that all frequences $\Omega_i^2$ are connected straightly with
the properties of the initial TCF
$a(t)$ only. The latter can be easily derived directly from the
experimental data \cite{Evans}-\cite{Yul}. Thus the system of
Eqs. \Ref{f57} has considerable utility for the experimental
investigations of statistical memory effects and non-Markov
processes in complex systems.

Among them it seems to us that one could propose more physical
interpretation of the different terms in the right side of the
three Eqs. (68). For example, term $-i\Delta A/ \Delta t$ is like a
dissipation, $\Delta^2 A/ \Delta t^2$ is like a inertia and
$\Omega^2 A(t)$ is like a
restoring force. Third derivative $\Delta^3 A/ \Delta t^3$  is the
finite-difference generic form of the Abraham-Lorenz force
corresponding to dissipation feedback due to radiative losses
[see for instance formula (3) in \cite{Johansen} for a recent
experimental evidence in frictional systems].

\section{Shannon entropy for the time correlation and time
memory in complex systems}
According to the results in section VI, the information measure for
the description of random processes in complex systems can
be expressed not only via TCF, but also by means of the certain set of time
memory functions. To accomplish that let us return to section III
in which we presented the geometrical picture of stochastic dynamics of 
correlation . In a line with Shannon \cite{Shannon} in
case of discrete source of information we were able to
determine a definite rate of generating information, namely the
entropy of the underlying stochastic information by introduction
fidelity evaluation function $\nu (P(x,y))$. Here the function
$P(x,y)$ is the two-dimensional distribution
of random variables $(x,y)$ and
\be
  \nu(P(x,y))=\int \int dxdy P(x,y)\rho(x,y),  \label{f70}
\ee
 where the function $\rho(x,y)$ has the general nature of the "distance"
between $x$ and $y$. As pointed by Shannon [17] the
function $\rho (x,y)$ is not a "metric" in the strict sense,
however, since in general it does not satisfy either
$\rho(x,y)=\rho(y,x)$ or $\rho(x,y)+\rho(y,z)\geq \rho (x,z).$.
It measures how undesirable it is according to
our fidelity criterion \Ref{f70} to receive y when x transmitted.
According to Shannon \cite{Shannon} any evalution of fidelity must correspond 
mathematically to the operation of a simple ordering of systems 
by the transmission of a signals within the certain tolerance.
 According to Shannon \cite{Shannon} the following is simple example of fidelity
evaluation function
\be
  \nu(P(x,y))=<(x(t)-y(t))^2>.           \label{f71}
\ee

In our case it is convenient to consider the initial
vector $A_k^0(0)$ as a variable x and the final vector
$A_{m+k}^m(t)$ at time $t=m\tau$ for a variable y. 
The distance function $\rho(x,y)$ [17]
\be
\rho(x,y)=\frac{1}{T}\int_0^T dt\{x(t)-y(t)\}^2  \label{f72} 
\ee
is the most commonly used measure of fidelity.

Taking into accout Eqs.\Ref{f70}, \Ref{f72} and the results in Section 3  
as the fidelity function one can use the following function 
of geometrical distance
\be
  \nu(P({\bf A}_k^0(0),{\bf A}_{m+k}^m(t))=2 k \sigma^2\{1-a(t)\},   \label{f73}
\ee
where distance function is
\be              
\rho({\bf A}_k^0(0),{\bf A}_{m+k}^m(t))=R^2({\bf
A}_k^0(0),{\bf A}_{m+k}^m(t)).    \label{f74}
\ee
  According to \cite{Shannon} partial solution of the general maximizing problem
for determining the rate of generating information of a source can be 
given using Lagrange's method and considering the following 
functional
\be
\int \int\{P(x,y) log\frac{P(x,y)}{P(x)P(y)}+\mu
P(x,y)\rho(x,y)+\nu(x) P(x,y)\}dxdy,   \label{f76}
\ee
where the function $\nu(x)$ and $\mu$ are unknown.
The following equation for the conditional probability can be
obtained by variation on $P(x,y)$ 
\be
  P_y(x)=\frac{P(x,y)}{P(y)}=B(x)\exp\{-\lambda \rho(x,y)\}.    \label{f77}
\ee
 This shows that with best encoding the conditional probability of a
certain cause for various received $y$, $P_y(x)$ will decline exponentially with
the distance function $\rho(x,y)$ between values the $x$ and $y$ in problem.
Unknown constant $\lambda$ is defined by the required fidelity, and 
function $B(x)$ in the case of continuous variables obeys the
normalization condition
\be
  \int B(x)\exp\{-\lambda \rho(x,y)\}dx=1.      \label{f78}
\ee
Since the distance function $\rho(x,y)$ \Ref{f72} is dependent
only on the vectors difference $\rho(x,y)=\rho(x-y)$,   
we get a simple solution for the special
case $B(x)=\alpha$
\be
P_y(x)=\alpha\exp \{-\lambda \rho(x-y)\}=\alpha
\exp\{-c[1-a(t)]\}         \label{f79}
\ee
instead of Eq.\Ref{f77}.
Constants $\alpha$ and $\lambda$ result from the
corresponding normalizing condition and in accordance with the required
fidelity. From the physical point of view the basic value of
solution \Ref{f79} is directly related to the occurence of the TCF
$a(t)$. Therefore,the solution \Ref{f79} describes the
state of the system with certain level and scale of correlation.

Now let us employ Shannon's solution for continuous variables
\Ref{f77}, \Ref{f79} and pass
to simplified discrete two-level description of the system. Then
let us consider the conditional probability \Ref{f79} which describes the
state on time axis at the moment $t=m\tau$ as corresponding to the
creation of correlation. Whereas the other state at the fixed moment $t=m\tau$
which accounts for the state with the absence (annihilation) of
correlation will exist. Let us introduce two
probabilities (see Fig. 2), which will fit normalizing condition
\be
 P_1(t)+P_2(t)=1,~~P_1(t)=P_{cc}(t),~~P_2(t)=P_{ac}(t),~~P_{cc}(t)+P_{ac}(t)=1. \label{f80}
\ee
 In the case of two lewels Shannon entropy
\be
   S=-\sum_{i=1}^2 P_i \ln P_i     \label{f81}
\ee
 increases at full disorder and takes its limiting value
\be
   \lim_{t \to \infty} S=\lim_{t \to \infty}S(t)=\ln 2.    \label{f82}
\ee
To find unknown parameters $\alpha$ and $c$ 
 in two-level description (creation and annihilation of
correlation) in Eq. \Ref{f77} we should take into account 
normalization condition,
principle of entropy increase \Ref{f82} at $t\to \infty$ and
of entropy extremality (presence of minimum)  at full order when
the following relationship: $\lim_{t \to o}a(t)=1$ is
true for the TCF. We obtained the following equation 
\bn
\lim_{t \to 0} S(t)=-\{\alpha \ln
\alpha+(1-\alpha)\ln(1-\alpha)\}=0 \nonumber
\en
for the parameters $\alpha$ and $c$ $(c
\geq 0,~0 \leq \alpha \leq 1)$ having regard to these requirements.
Among  two solutions $(\alpha_1=1,~\alpha_2=0)$ only the first 
one ($\alpha_1=1$) has physical sense. Two probabilities calculated 
by means of Eq.\Ref{f79} will satisfy conditions
\Ref{f80}, \Ref{f82}
\be
 P_1(t)=P_{cc}(t)=\exp\{-\ln2[1-a(t)]\},   
 \label{f83}
\ee
\be
  P_2(t)=P_{ac}(t)=1-\exp\{ -\ln2[1-a(t)]\}.  
  \label{f84}
\ee
respectively. In acordance with two -lewel description it would
be convenient to deal with two dynamic channels of entropy (creation (cc)
and annihilation (ac)) of correlation (see Fig.2)
\be
 S_{cc}(t)=\ln2\{ 1-a(t)\} \exp\{ -\ln2[1-a(t)]\},    \label{f85}
\ee
\be
  S_{ac}(t)=-\{1-\exp[(-\ln2(1-a(t))]\} \ln \{1-\exp[(-\ln2(1-a(t))]\}.  \label{f86}
\ee
The probabilities obtained are in the line with full dynamic (time
dependent) information Shannon entropy 
\bn
S_0(t)=S_{cc}(t)+S_{ac}(t)=\ln2\{ 1-a(t)\} \exp\{
-\ln2[1-a(t)]\} \nonumber \\-\{1-\exp[(-\ln2(1-a(t))]\} \ln
\{1-\exp[(-\ln2(1-a(t))]\}.  \label{f87}
\en
The entropy introduced in to Eqs. \Ref{f85}-\Ref{f87} 
characterized a quantitative measure of disorder in the system related
to creation and annihilation of dynamic correlation. The
probabilistic and entropy channels \Ref{f83}-\Ref{f86} possess
the following asymptotic behavior 

if $a(t)\to 1,$
\bn
P_1(t)&=&P_{cc}(t)\cong 1+\ln2[a(t)-1],
P_2(t)=P_{ac}(t)\cong \ln2[1-a(t)],\nonumber \\
S_{cc}(t)&\cong&-\{1+\ln2[a(t)-1]\}\ln\{1+\ln2[a(t)-1]\},
S_{ac}(t)\cong-\ln2[1-a(t)]\ln\{\ln2[1-a(t)]\}, \nonumber \\
S_0(t)&\cong&-\{1+\ln2[a(t)-1]\}\ln\{1+\ln2[a(t)-1]\}-
\ln2[1-a(t)]\ln\{\ln2[1-a(t)]\} \nonumber;
\en
and if $a(t)\to 0,$
\bn
P_1(t)&=&P_{cc}(t)\cong\frac{1}{2}\{1+\ln2\cdot a(t)\},~P_2(t)=P_{ac}(t)
\cong \frac{1}{2}\{1-\ln2\cdot a(t)\},\nonumber \\
S_{cc}(t)&\cong&-\frac{1}{2}\{1+\ln2\cdot a(t)\}\ln\{\frac{1}{2}[1+\ln2\cdot
a(t)]\},~S_{ac}(t)\cong-\frac{1}{2}\{1-\ln2\cdot
a(t)\}\ln\{\frac{1}{2}[1-\ln2 \cdot 
a(t)]\},\nonumber \\
S_0(t)&\cong&-\frac{1}{2}\{1+\ln2\cdot
a(t)\}\ln\{\frac{1}{2}[1+\ln2 \cdot 
a(t)]\}-\frac{1}{2}\{1-\ln2\cdot
a(t)\}\ln\{\frac{1}{2}[1-\ln2 \cdot
a(t)]\}. \nonumber \\
\en
Subsequent boundary conditions result from the equations above in
terms of Bogolubov's principle of attenuation of correlation
\Ref{f7} as follows:
\be
 \lim_{t \to 0}P_{cc}=1,~~\lim_{t \to 0} P_{ac}(t)=0,     \label{f88}
\ee
\be
\lim_{t \to 0}S_{cc}(t)=0,~~\lim_{t \to 0}S_{ac}(t)=0,~~\lim_{t
\to 0}S_0(t)=0,     \label{f89}
\ee
\be
\lim_{t \to \infty}P_{cc}(t)=\frac{1}{2},~\lim_{t \to
\infty}P_{ac}=\frac{1}{2}, \lim_{t \to \infty}S_{cc}(t)=\lim_{t
\to \infty}S_{ac}(t)=\frac{\ln2}{2},~\lim_{t \to
\infty}S_0(t)=\ln2.  \label{f90} 
\ee
It is conditions \Ref{f88} that give us an opportunity to
present two different states associated with the creation
(cc)(in the time moment $t=0, \ P_{cc}=1$) and annihilation (ac)
(at the point $t=0,~P_{ac}(0)=0$) of correlation. Owing to
discretness of the TCF $a(t)$ all functions 
$P_{\alpha \beta},~S_{\alpha \beta}$ as well as $S_0(t)$ $(\alpha=a,c;~
\beta=c)$ are discrete in the real complex systems.

The results obtained in Section VI permit us to present the set of
entropies for the states connected with
 the set of orthogonal variables $W_i$ and set of memory functions
$ M_i(t)=\{M_1(t),M_2(t),M_3(t),\cdots\}$. On the analogy to
Eqs.\Ref{f83}-\Ref{f90}
these functions describe non-Markov and memory effects in the
 system under discussion
\be
 P_1^{M_i}(t)=P_{cM_i}(t)=\exp\{ -\ln2[1-M_i(t)]\},    \label{f91}
\ee
\be
 P_2^{M_i}(t)=P_{am_i}(t)=1-\exp\{ -\ln2[1-M_i(t)]\},       \label{f92}
\ee
\bn
S_i(t)=\ln2[1-M_i(t)] \exp\{\ln2[1-M_i(t)|\}- \nonumber \\
\{ 1-\exp[-\ln2(1-M_i)]\}\ln\{ 1-\exp[-\ln2(1-M_i)]\},        \label{f93}
\en
where $i=1,2,3$.
Four corresponding entropies $S_0 (t),~S_1(t),S_2(t)$ and $S_3(t)$
and their power frequency spectra
are available from the set of four time functions (TCF $a(t)$ and
three memory functions $M_1(t),~M_2(t),~M_3(t)$). Equations
\Ref{f83}-\Ref{f93} are of great value because
they allow us to estimate stochastic
dynamics of the real complex systems with discrete time. As a
matter of principle the first three
memory functions $M_i(t)~(i=1,2,3)$ are easy
to find via Eqs. \Ref{f57}. Using dimensionless parameter 
$\varepsilon_1=\tau^2\Omega_1^2$
and solution of the first finite-difference Eq.\Ref{f57}we can
calculate the discrete function $M_1(j\tau)$ at 
the points $j=0,1,2,\cdots$ as follows:
\bn
M_1(0)=1,~M_1(\tau)=-a(2 \tau)+\varepsilon_1^{-1}\{a(2\tau)-a(3\tau)\},
\nonumber \\
M_1(2\tau)=-\{ a(2 \tau) M_1(\tau)+a(3\tau)\}+\varepsilon_1^{-1}\{
a(3\tau)-a(4\tau)\},\nonumber \\
M_1(3\tau)=-\{a(2\tau) M_1(2\tau)+a(3\tau)M_1(\tau)+a(4\tau)M_1(0)\}+
\varepsilon_1^{-1}\{ a(4\tau)-a(5\tau)\},\nonumber \\
...................................................................
\nonumber \\  
M_1(m\tau)=-\sum_{j=0}^{m-1}M_1(j\tau)a\{(m+1-j)\tau\}+
\varepsilon_1^{-1}[a\{ (m+1)\tau\}-a\{ (m+2)\tau\}].      \label{f94}
\en
In general case solving the chain of Eqs. \Ref{f55},\Ref{f57} we
can found the recurrence relations between the memory
functions of junior and higher orders in the following form
\be
M_s(m\tau)=-\sum_{j=0}^{m-1}
M_s(j\tau)M_{s-1}((m+1-j)\tau)+\varepsilon_s^{-1}\{
M_{s-1}((m+1)\tau)-M_{s-1}((m+2) \tau)\},~~\varepsilon_s=\tau^2\Omega_s^2,~s=1,2,3,\cdots     \label{f95}
\ee
The relations obtained  allow us to derive straightly the necessary memory
functions $M_s(t)$ of any order $s=1,2,...$ from
experimental data using the registered TCF $a(m\tau)$
\cite{Coffey}, \cite{Yul}. Relaxation
frequencies $\Omega_i^2$, $i=1,2,3,...$, given in Eqs.
\Ref{f95} are available to experimental registration. Thus,
it is fair to say that the applications of Eqs.\Ref{f95} will
open up fresh opportunities for detailed study of statistical
properties of correlations in the complex systems. The very fact of
existence of finite -difference Eqs. \Ref{f55}, \Ref{f57}
enables us to develop any functions directly from the experiment.
Therefore, the availability of discretness permits to
enhance substantially the capability to get information for the
complex systems' state.

In conclusion let us show the equations, which characterize the rate
of entropy production. It is obvious from conditions
\Ref{f88}-\Ref{f90} as well as  Eqs. \Ref{f83}-\Ref{f87} that the rate
of entropy growth $\frac{\partial S}{\partial t}$ 
within the interval $(0,\infty)$ takes different sign values 
and is determined by the entropy behavior in the channels
of creation and annihilation of correlation 
\be
\frac{\partial S_0}{\partial t}=\left(\frac{\partial
S_1^{(0)}}{\partial t}\right)+\left(\frac{\partial
S_2^{(0)}}{\partial t}\right)=\left(\frac{\partial
S_{cc}(t)}{\partial t}\right)+\left(\frac{\partial
S_{ac}(t)}{\partial t}\right),      \label{f96}
\ee
\be
\frac{\partial S_1^{(0)}(t)}{\partial t}=-\ln2 \cdot a'(t) \exp\{
-\ln2[1-a(t)]\} \{1-\ln2[1-a(t)]\},     \label{f97}
\ee
\be
\frac{\partial S_2^{(0)}(t)}{\partial t}=-\ln2a'(t) \exp\{
-\ln2[1-a(t)]\} \{1+\ln[1-\exp[-\ln2(1-a(t)]\},    \label{f98}
\ee        
\be
\frac{\partial S_0(t)}{\partial t}=\ln2a'(t) \exp\{
-\ln2[1-a(t)]\}\{\ln\{1-\exp[-\ln2(1-a(t))]\}+\ln2[1-a(t)].    \label{f99}
\ee
The derivatives $a'(t)$ and $S_0'(t)$ here should be read 
in terms of Eqs. \Ref{f8}, \Ref{f13}.Since the
derivative $a'(t)$ is finite within the whole time interval
$(0,\infty)$:  $|a'(t)|<c$, (where
$c$ is positive constant) the rate of entropy growth obeys the
following boundary conditions
\be
\lim_{t \to 0}\left(\frac{\partial S_0}{\partial
t}\right)=0, ~~\lim_{t \to \infty}\left(\frac{\partial S_0}{\partial
t}\right)=0.    \label{f100}
\ee
Formulas \Ref{f96}-\Ref{f100} are useful for the discussion of the
experimental data. Close inspection of these equations shows
that the behaviour of derivative $\left(\frac{\partial S_0}{\partial
t}\right) $ is described in many respects by the function
$a'(t)=\tau^{-1}[a(t+\tau)-a(t)$], which is in its turn can be
obtained from the time series observed. Relations analogous to
 Eqs. \Ref{f96}-\Ref{f100} are easily available for the
sequence of memory functions $M_i(t)$ (55) as well.
                   
\section{Application on analysis of stochastic dynamics of R-R intervals 
    in human ECG's}

Let us use the stochastic dynamics of RR-intervals from human ECG's
to illustrate of some practical value of the approach developed. It is well 
known \cite{Zebrowski}, \cite{Webber}, \cite{Yulm}-\cite{Peng}
 that the statistical analysis of related dynamics allows 
the reliable quantitative characteristics of the human
cardiovascular system states and trusty diagnostics of the
various heart diseases \cite{Gold}-\cite{Guil}.

Most investigators into heart rate dynamics have emphasized
continuous functions, whereas the heart beat itself is in a
crucial respect a discrete event. We present here experimental
evidence that by considering this quality, the behavior of
RR intervals may be appreciated as a result of discrete
dynamics. 
To demonstrate effectiveness of non-Markovian approach we only
take four typical particular cases from the whole the set of
experimental data \cite{ssyul}, which are available at our disposal. They are
related to the case of healthy man (a), patient with a rhythm 
driver migration (b), patient after
myocardial infarction (MI) (c), and patient after myocardial
infarction (MI) with subsequent sudden cardial death (SCD) (d).
Following standard medical practice, each from 112 person had an
age, sex, and disease status matched pair serving as the
control. 

Results of our calculations, based on formulas of the theory and
presented in previous sections, are shown on Figs. 3-8.
It is necessary to mark, that as a matter of convenience all variables and 
functions in a Figs. 3-8 are submitted in dimensionless form. 
Frequency $\omega$ everywhere 
is indicated in terms of units of $2 \pi / \tau$. The orthogonal variables 
$W_0$ and $W_2$ in a Figure 3 are written 
in units of $\tau$ and $\tau^{-1}$, respectively. Frequency
spectra $\mu_0(\omega)$, $\mu_1(\omega)$ and $\mu_2(\omega)$ in  Figs. 4-6 are figured 
in terms of units of $\tau^2$. Values $\epsilon_1(\omega)$ and
$\epsilon_2(\omega)$ in  Figs. 7, 8 are dimensionless values.
Fig.3 shows phase trajectories, obtained for four different groups of
patients in orthogonal variable's $(W_0, W_2)$ plane. Let us
remind ourselves, that in correspondence with formulas
\Ref{f64}, \Ref{f68} the variable $W_0$  presents RR-intervals fluctuations, and 
$W_2$ is the second orthogonal variable and due to Eq. \Ref{f68}
is combination of an inertia force minus a restoring force. 
These variables have dimensions $\tau$ and $\tau^{-1}$ respectively,
where $\tau=<\l_{RR}>$ is the average value of
RR-interval in time sequence. Set of characteristic parameters is collected in
Table I. Let us mention the strong difference of numerical value
of the first general relaxation frequency $\Omega_1$ frequency 
for four different groups of patients.
Figures. 4-6 show power frequency spectra for three different time
functions for typical patients from four different groups. Fig.
3(a) corresponds to a strange attractor, fig. 3(b) corresponds to
quasi-periodic motion,fig.3(c) and fig.3(d) demonstrate the
obviously expressed correlation of phase variables $W_0$ and
$W_2$. Although the
frequency $\omega$ is measured in units of $\frac{2\pi}{\tau}$ and power
in $\tau^2$, respectively. Figure 4 shows power spectrum of TCF
fluctuations of RR-intervals. The data, shown on Figs. 5, 6 are
correspondingly related to power spectra of first and second
memory functions. Functions themselves are calculated from
formulas \Ref{f57}, \Ref{f68} and \Ref{f95}.

Figures. 7, 8 require special explanation. They show frequency
spectra of first two points $\epsilon_1(\omega)$ and
$\epsilon_2(\omega)$ of statistical spectra of non-Markovity
parameter (NMP) ${\epsilon_i}$, where $i=1, 2,...$ . A presentation
of NMP spectrum was introduced earlier in Refs. \cite{Shur},
\cite{Shury}  and was then
used in statistical physics of liquids \cite{Ren},\cite{Rena}.
Close to that given in  
\cite{Shur}, \cite{Shury} definitions of non-Markovity were
developed later in Refs. \cite{Brin}-\cite{Teichl}. In
comparision with  Refs. \cite{Shur}-\cite{Ren} here 
we generalize NMP conception for frequency dependent case
\bn
\epsilon_i(\omega)=\left \{ \frac{\mu_{i-1}(\omega)}
{\mu_i(\omega)} \right \} ^{\frac{1}{2}},
\nonumber
\en
where $i=1,2,..$ and $\mu_i(\omega)$ is power frequency spectrum
of i-th level. 

As is shown by Yulmetyev et al in articles \cite{Shur}-\cite{Rena}
 NMP value of $\epsilon_i$ allows to obtain quantitative
estimate of non-Markovity effects and statistical collective
memory in random changes of experimentally measured data.

Parameter $\epsilon_i$ allows to divide
all processes in three important cases \cite{Shur}-\cite{Rena}.
 Markovian processes correspond to
$\epsilon\gg1$, while quasimarkovian processes correspond to
situation with $\epsilon>1$. The limit case $\epsilon
\sim 1$ describes non-Markovian processes. In this case the time scale of
memory processes and correlations (or junior and senior memory
functions) coincide with each other.

From Figs. 3-8 one can easily obtain sharp
differences between four groups of patients for all types of
frequency spectra. For instance, frequency spectrum of TCF power
for healthy [Fig. 4a] is almost reproduced in NMP
$\epsilon_1(\omega)$ spectrum given in Fig. 7a. Also it is
slightly deformed in the spectra of first [Fig. 5a] and
second [Fig. 6a] memory functions and is strongly
transformed in NMP $\epsilon_2(\omega)$ spectrum (Fig. 8a).
Sharp peak in the vicinity of the point with $\omega \sim 0.125
f.u$, being characteristical for the patient (b), is seen
in the power spectrum of first and second MF's (Fig. 5(b), 6(b)).
However for other spectra of type b (for example Figs. 6b, 7b, 8b)
quite complicated structure appears. Frequency spectrum of type
(c), which is characteristic for IM, contains two sharply
expressed spectral peaks nearly the frequencies, approximately 0.2 and
0.4 f.u. on the background of low intensity  white noise  .
These peaks are conserved in the spectra of first (Fig. 5c) and
second (Fig. 6c) MF. In NMP spectra $\epsilon_i(\omega), \
\epsilon_2(\omega)$, complicated structure of spectral lines also
appears. In characteristic case of patient with SCD
frequency spectra of type (d) everywhere contain sharp peaks
close to frequency 0.25 f.u. We would like to mention that all
frequency spectra (5,6,7 and 8) are persuasively 
saying for strongly expressed non-Markovity for time change of
RR- intervals.

Figs 7a-d and 8a-d shows, that all values of NMP
$\epsilon_1(\omega)$ and $\epsilon_2(\omega)$ lie in small
interval of values (0 $\div$ 30).

This fact convincingly tell us about characteristic statistic
memory and noticeable non-Markovity effects in statistical
dynamics of RR-intervals from human ECG's. Obtained results on
non-Markovian properties of temporal behavior of RR-intervals
justify significant and characteristic differences in data for
all 4 groups of patients.
We hope that the use of non-Markovian dynamics in the spirit
of developed theory will incorporate development of more precise
estimate of the state of cardiovascular systems for healthy as
well as for more careful diagnostics of different patients.

\section{Discussion}
The present paper deals with two interrelated important results. 
The first one is connected with the establishment of the chain of
finite-difference non-Markov kinetic equations for the discrete
TCF. In this case the state of complex systems
at the definite level of correlation is described by two vectors constructed
over the strict determined rules. It is natural finite-difference
equation of motion, being the peculiar analog of Liouville
equations for the initial dynamic variables, which are of
particular interest for
our analysis. In the subsequent discussion we employ the
strict deduced mathematical fact of the existence of the normalized
TCF. Due to the operation of
scalar  product the availability of TCF makes it possible to
introduce the projection operators in the space of vectors of states. Those
projection operations and matrix elements of Liouville's quasioperator
ensure the splitting of natural equations of motion and then
they are solved in the closed finite-difference form. 
Using Gram-Schmidt orthogonalization procedure we  
find an infinite set of the orthogonal dynamic random variables.
This it allows us to obtain the whole infinite chain of
finite-difference kinetic equations for the initial discrete
TCF. These equations contain the set of all memory functions
characterizing the complete spectrum of non-Markov processes and
statistical memory effects in the complex system. The presence of
discretness and the very fact of the existence of
finite-difference structure enable, in principle,  to find 
all memory functions solving successively kinetic equations for
the TCF.
Parameters of these equations can  be easily obtained from the
experimentally registered TCF. In chaotic dynamics of complex
systems the TCF above plays the role  
similar to that of the statistical integral in equilibrium
statistical physics. 

Another important result of our work is the dynamic (time
dependent) information Shannon entropy given in terms of the TCF. It allows
us to use the information measure for the quantitative characteristic
of two interrelated correlation channels. One of them corresponds
to the creation of time correlation and the other to the
annihilation of correlation.

For that as we employ one of the classical Shannon's
results \cite{Shannon}, related to the introduction of fidelity
evalution function and distance function between two vectors of
state. The existence of a new information measure opens up new
fields for exploration of information characteristics
of complex systems. In particular, some interesting data arise from 
calculations frequency spectra of power of information entropy. 

The important consequence of the results obtained is the usage  of
power spectra of memory functions
$M_j(m\tau)$, where $m=0,1,2,3,...$ and $j=1,2,3,...$.
The set of three junior memory functions with numbers $j=1,2,3$~
 provides the basis for the pseudohydrodynamical description of the
complex system. In practice, any memory function can be
extracted from the experimental
time sets and experimentally recorded TCF. These criteria
provide the possibility to get reliable information about
non-Markov processes and memory effects in natural evolution of
complex systems. In principle, the new point in the analysis of
complex systems arises from the opportunity to construct
the dynamical information Shannon entropy for the
experimental memory functions. Undoubtedly, detection of the 
frequency spectra of power of entropy for memory functions gives
us new unique information about the statistical non-Markov 
properties as well as memory effects in complex systems of various nature.

Application of the theory developed on the analysis of dynamics
of RRintervals from human ECG's strongly suggest the
substantially non-Markovian properties of the this dynamics.
Here we have obtained non-Markovian quantitative characteristics
for the fourth various groups of patients. One might expect
this method may be use in distinguishing healthy from pathalogic
data sets based in differences in its non-Markovian properties.

In conclusion it may be said that this paper describes a
first-principle derivation of a hierarchy of finite-difference
equations for time correlation function of out-of-equilibrium
systems without Hamiltonian. The approach developed seems to
have potentials and offer few advantages over the usual
Hamiltonian point of view. A similar situation are true
apparently with regard to turbulence, aging, for istance, as in
spin glasses and glasses as well as experimental time series for
living, social and natural complex systems (physiology,
cardiology, finance, psychology, and seismology, etc.) 

By way of illustration it is significant that the anomalous
scaling of simultaneous correlation function in turbulence is
intimately related to the breaking of temporal scale invariance,
which is equivalent to the appearence of infinitely many time
times scales in the time dependence of time-correlation
functions. In Refs. \cite{Dae} was addresed temporal
multiscaling on the basis of the continued fraction
representation of turbulent correlation function \cite{Gros}
within the framework the Zwanzig-Mori formalism \cite{Zwanzig},
\cite{Mori} which was applied to the time correlation function
in turbulence. It has been shown by Grossman and Thomas
\cite{Gros} that the Zwanzig-Mori formalism applied to turbulent
systems described by Navier- Stokes-like equations.

Mode coupling equatins have been considered in various areas of
many particle physics for an approximate treatment of the
dynamics of particles in glasses \cite{Haus}, \cite{Gof}. These
equations are obtained if one represent within the Zwanzig-Mori
formalism \cite{Zwanzig}, \cite{Mori} correlation functions in
terms of memory kernels and then expressed the latter via a
factorization approximation in terms of the former for the glass
transition of molecular liquids \cite{Schil}. It was been found
by Heuer et al. \cite{Heuer} that a model-free interpretation of
higer-order correlation function determined by NMR reveals
important information about the complex dynamics close to glass
transition of polymers.
This is now been demonstrated with spin glasses \cite{Hoffm}
that how a hierarchical model of spin glasses relaxation can
display aging behavior in the time scale, similarly to what is
found in spin glasses and other complex systems out of
thermodynamical equilibrium. 

The application of the approach
developed on the analysis of the temporal behavior of complex
systems of various nature will be available in our forthcoming
papers.

\section{Acknowledgments}
RMY wishes to thank the DAAD for support and
Lehrstuhl f\"ur Theoretische Physik,Institute of Physics 
at Augsburg University for hospitality. This work was partially supported
by Competetive Center for Fundamental Research at St-
Peterburg University (grant N. 97-0-14.0-12), Russian Humanitar
Science Fund (grant N.00-06-00005a) and NIOKR RT foundation (grant
N. 14-78/2000(f)). Authors acknowleges Dr. I. Goychyk for valuable
criticism, Prof. I.A.Latfullin and
M.Dr.G.P. Ischmurzin (Department of Therapy, Kazan State 
Medical University) for the
presentation and discussion of hyman short-time ECG's data.
 
\newpage
\section{Figure captions}
Fig.1. Simple geometrical notion on vectors, their scalar
product and normalized TCF of random variables.

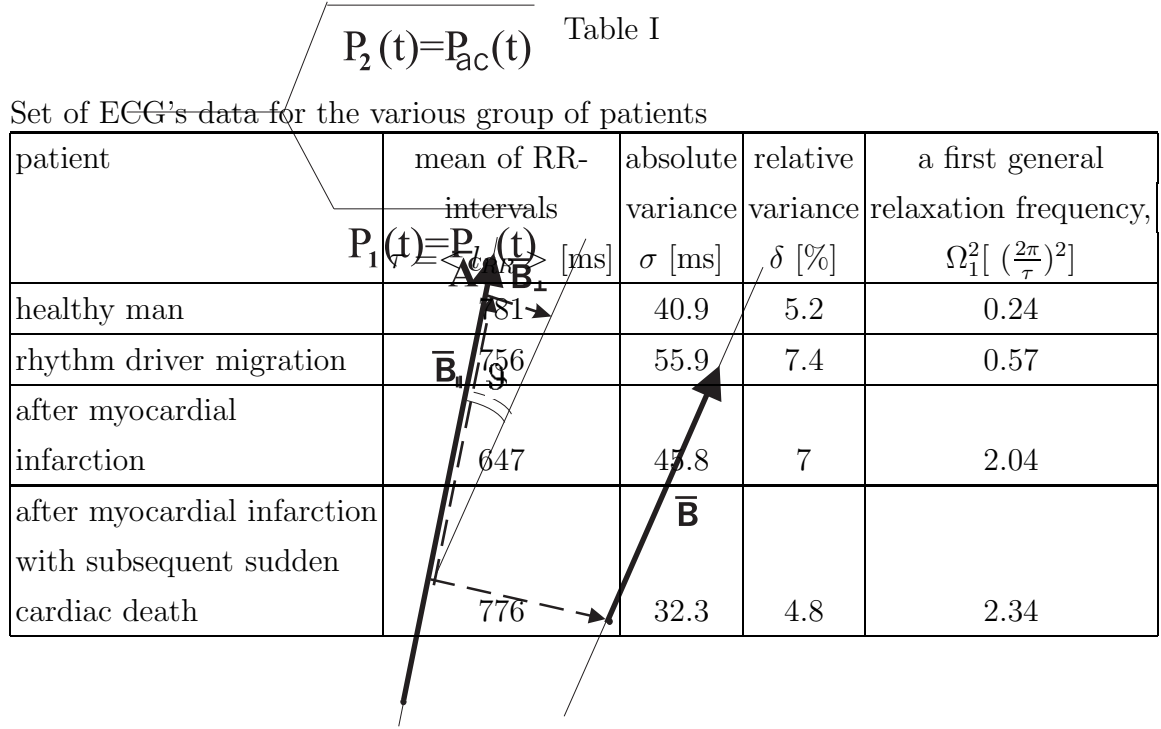
Fig.2. Scheme of simplified two-lewel description of a complex
systems state. Two probability $P_{cc}(t)$ and $P_{ac}(t)$
decsribes a stochastic processes of creation (existence) and
annihilation (decay) of time correlation.

Fig 3. Phase-time portreit in orthogonal variables ($W_0, \
W_2$) plain (see
formulas \Ref{f66}, \Ref{f68} for fourth group of patients:
healthy (a), patient with rhythm driver migration (b), patient
after myocardial infarction (c) , and patient after MI with
subsequent SCD (d). As a matter of fact us utilised dimensionless variables $W_0/
\tau$ and $W_2/ \tau^{-1}$.

Fig 4. Frequency spectrum of power $\mu_0(\omega)$ for TCF of 
fluctuation of RR-
intervals for fourth patient groups: healthy (a),
 patient with rhythm driver migration (b), patient
after myocardial infarction (c) , and patient after MI with
subsequent SCD (d). The schedule is submitted in dimensionless units. The frequency is marked in 
terms of units of ($2 \pi/ \tau$), the function 
$\mu_0(\omega)$ is figured in units of $\tau^2$.

Fig 5. Frequency spectrum of power $\mu_1(\omega)$ for the first MF $M_1(t)$ for
fourth patient groups:healthy (a), patient with rhythm driver migration (b), patient
after myocardial infarction (c) , and patient after MI with
subsequent SCD (d). The schedule is submitted in dimensionless units. The frequency is marked in 
terms of units of ($2 \pi/ \tau$) , the function $\mu_1(\omega)$  is
figured in units of $\tau^2$. 

Fig 6.  Frequency spectrum of power $\mu_2(\omega)$ for the second MF $M_2(t)$ for
fourth patient groups:healthy (a), patient with rhythm driver migration (b), patient
after myocardial infarction (c) , and patient after MI with
subsequent SCD (d). The schedule is submitted in dimensionless units. The frequency is marked in 
terms of  units of ($2 \pi/ \tau$) , the function $\mu_2(\omega)$  is
figured in units of $\tau^2$.

Fig 7. Frequency spectrum of the first point in the statistical
spectrum on non-Markovity parameter $\epsilon_1(\omega)$ for
fourth patient groups:healthy (a), patient with rhythm driver migration (b), patient
after myocardial infarction (c) , and patient after MI with
subsequent SCD (d). The schedule is submitted in dimensionless units. The 
frequency is marked in  terms of units of $\tau^2$ .

Fig 8. Frequency spectrum of the second point in the statistical
spectrum on non-Markovity parameter $\epsilon_2(\omega)$ for
fourth patient groups:healthy (a), patient with rhythm driver migration (b), patient
after myocardial infarction (c) , and patient after MI with
subsequent SCD (d). The schedule is submitted in dimensionless units. The frequency is marked in 
terms of  units of $\tau^2$.

\begin{thebibliography}{10}
\bibitem{ssyul}\r{The R-R recordings were drawn from
Division of Cardiology of Emergency Resque (Kazan, Tatarstan)
congestive heart-failure database comprising 7 records from
normal patients (age: 26-59 years; mean 45 years) and 112
records from severe congestive heart disease patients (age:
25-69 years, mean 53 years). The recordings, which form a
standard database for evaluting the merits of various measures
for identificatying heart disease, were made with a Holter
monitor, digitized at a fixed value of 128 Hz. We have used the
short-time series of RR- intervals.}
\end{thebibliography}
\end{document}